\documentclass[11pt]{article}
\usepackage{axodraw}
\usepackage{epsfig}
\usepackage{graphicx}
\usepackage{amsfonts}
\usepackage{amsmath}
\usepackage{bbm}
\allowdisplaybreaks[1]

\input pix.sty

\def\undertilde#1{\mathop{\vtop{\ialign{##\cr$\textstyle{#1}$\cr%
\noalign{\kern1pt\nointerlineskip}\hfil$\mathchar"0365$\hfil\cr}}}}
\def\wideundertilde#1{\mathop{\vtop{\ialign{##\cr$\textstyle{#1}$\cr%
\noalign{\kern1pt\nointerlineskip}\hfil$\mathchar"0367$\hfil\cr}}}}

\renewcommand{\eq}{eq.~}
\renewcommand{\eqs}{eqs.~}
\renewcommand{\se}{sec.~}

\renewcommand{\fig}{fig.~}
\renewcommand{\figs}{figs.~}

\newcommand{\ko}{k^{ }_0}
\newcommand{\CF}{C_\rmii{F}}
\newcommand{\Nf}{N_{\rm f}}

\newcommand{\Nc}{N_{\rm c}}

\newcommand{\mE}{m_\rmii{E}}

\newcommand{\gammaE}{\gamma_\rmii{E}}

\newcommand{\rmO}{{\mathcal{O}}}

\renewcommand{\bG}{\beta_\rmi{G}}

\def\lsi{\raise0.3ex\hbox{$<$\kern-0.75em\raise-1.1ex\hbox{$\sim$}}}
\def\gsi{\raise0.3ex\hbox{$>$\kern-0.75em\raise-1.1ex\hbox{$\sim$}}}
\newcommand{\lsim}{\mathop{\lsi}}
\newcommand{\gsim}{\mathop{\gsi}}

\newcommand{\nB}{n_\rmii{B}}

 \renewcommand{\nB}[1]{n_\rmii{B{#1}}}
\newcommand{\rmii}[1]{{\mbox{\tiny\rm{#1}}}}
\newcommand{\rmiii}[1]{{\mbox{\tiny{$\scriptstyle{\rm#1}$}}}}
\newcommand{\re}{\mathop{\mbox{Re}}}
\newcommand{\im}{\mathop{\mbox{Im}}}

\newcommand{\Tint}[1]{{\hbox{$\sum$}\!\!\!\!\!\!\!\int\,}_{\!\!\!\!\raise-0.9ex\hbox{$\scriptstyle{#1}$}}}
\newcommand{\Tinti}[1]{{{\Sigma}\!\!\!\!\raise0.3ex\hbox{$\int$}_\rmii{${#1}$}}}

\newcommand{\bi}{\begin{itemize}}
\newcommand{\ei}{\end{itemize}}

\newcommand{\hide}[1]{ }

\def\scfc{1.4}  
\renewcommand{\picc}[1]{\PIC{#1}{\pwcc}{\phgt}{\scfc}}
\def\ScatA{\picc{%
 \Line(20,5)(50,5)%
 \Line(10,5)(20,5)%
 \Line(50,4.7)(50,25.3)%
 \Line(10,25)(40,25)%
 \Line(40,25)(50,25)%
 \Line(10,4.7)(10,25.3)%
 \Lgl(30,5)(30,25)%
}}
\def\ScatB{\picc{%
 \Line(10,5)(50,5)%
 \Line(50,4.7)(50,25.3)%
 \Line(10,25)(50,25)%
 \Line(10,4.7)(10,25.3)%
 \Agl(30,25)(6,180,360)%
}}
\def\ScatC{\picc{%
 \Line(10,5)(50,5)%
 \Line(50,4.7)(50,25.3)%
 \Line(10,25)(50,25)%
 \Line(10,4.7)(10,25.3)%
 \Agl(30,5)(6,0,180)%
}}
\def\ScatD{\picc{%
 \Line(10,5)(50,5)%
 \Line(50,4.7)(50,25.3)%
 \Line(10,25)(50,25)%
 \Line(10,4.7)(10,17)%
 \Line(10,17)(10,25.3)%
 \Agl(10,25)(8,270,360)%
}}
\def\ScatE{\picc{%
 \Line(10,5)(50,5)%
 \Line(50,4.7)(50,25.3)%
 \Line(10,25)(50,25)%
 \Line(10,15)(10,25.3)%
 \Line(10,4.7)(10,15)%
 \Agl(10,5)(8,0,90)%
}}
\def\ScatF{\picc{%
 \Line(10,5)(50,5)%
 \Line(50,4.7)(50,15)%
 \Line(50,15)(50,25.3)%
 \Line(10,25)(50,25)%
 \Line(10,15)(10,25.3)%
 \Line(10,4.7)(10,15)%
 \Lgl(10,15)(50,15)%
}}
\def\ScatG{\picc{%
 \Line(10,5)(50,5)%
 \Line(50,4.7)(50,25.3)%
 \Line(10,25)(50,25)%
 \Line(10,4.7)(10,25.3)%
 \Agl(10,15)(6,270,90)%
}}
\def\ScatH{\picc{%
 \Line(10,5)(50,5)%
 \Line(50,4.7)(50,25.3)%
 \Line(10,25)(50,25)%
 \Line(10,4.7)(10,25.3)%
 \Agl(50,15)(6,90,270)%
}}
\def\ScatI{\picc{%
 \Line(10,5)(50,5)%
 \Line(50,4.7)(50,17)%
 \Line(10,25)(50,25)%
 \Line(10,4.7)(10,25.3)%
 \Line(50,17)(50,25.3)%
 \Agl(50,25)(8,180,270)%
}}
\def\ScatJ{\picc{%
 \Line(10,5)(50,5)%
 \Line(50,13)(50,25.3)%
 \Line(10,25)(50,25)%
 \Line(10,4.7)(10,25.3)%
 \Line(50,4.7)(50,13)%
 \Agl(50,5)(8,90,180)%
}}
\newcommand{\picmin}[1]{\PIC{#1}{\pwcc}{\phgt}{0.5}}
\def\ScatLegend{\picmin{%
 \LongArrow(10,10)(50,10)%
 \LongArrow(10,10)(10,50)%
 \Text(30,10)[c]{$r_\perp$}%
 \Text(12,28)[c]{$\tau$}
}}
\makeatletter \@addtoreset{equation}{section} \makeatother
\renewcommand{\theequation}{\arabic{section}.\arabic{equation}}
\makeatletter
\renewcommand\section{\@startsection {section}{1}{\z@}%
                                   {-5.5ex \@plus -1ex \@minus -.2ex}
                                   {2.3ex \@plus.2ex}%
                                   {\normalfont\large\bfseries}}
\renewcommand\subsection{\@startsection{subsection}{2}{\z@}%
                                     {-3.25ex\@plus -1ex \@minus -.2ex}%
                                     {1.5ex \@plus .2ex}%
                                     {\normalfont\normalsize\bfseries}}
\renewcommand\thesection {\@arabic\c@section}
\renewcommand\thesubsection   {\thesection.\@arabic\c@subsection}
\renewcommand{\@seccntformat}[1]{%
\csname the#1\endcsname.\hspace{1.0em}}
\makeatother

\begin{document}

\flushbottom

\begin{titlepage}
\begin{flushright}
\vspace*{1cm}
\end{flushright}
\begin{centering}
\vfill

{\Large{\bf
 Light-cone Wilson loop in classical lattice gauge theory
}} 

\vspace{0.8cm}

M.~Laine, 
A.~Rothkopf 

\vspace{0.8cm}


{\em
Institute for Theoretical Physics, 
Albert Einstein Center, University of Bern, \\
Sidlerstrasse 5, CH-3012 Bern, Switzerland
}

\vspace*{0.8cm}

\mbox{\bf Abstract}
 
\end{centering}

\vspace*{0.3cm}
 
\noindent
The transverse broadening of an energetic jet passing through a non-Abelian
plasma is believed to be described by the thermal expectation value of a
light-cone Wilson loop. In this exploratory study, we measure the light-cone
Wilson loop with classical lattice gauge theory simulations. We observe, 
as suggested by previous studies, that there are strong interactions 
already at short transverse distances, which may lead to more efficient jet
quenching than in leading-order perturbation theory. We also verify that the
asymptotics of the Wilson loop do not change qualitatively when crossing the
light cone, which supports arguments in the literature that infrared
contributions to jet quenching can be studied with dimensionally reduced
simulations in the space-like domain. Finally we speculate on possibilities
for full four-dimensional lattice studies of the same observable, perhaps by
employing shifted boundary conditions in order to simulate ensembles boosted
by an imaginary velocity.

\vfill

 
\vspace*{1cm}
  
\noindent
July 2013

\vfill

\end{titlepage}

%
\section{Introduction}
\la{se:intro}

When an energetic jet traverses a strongly interacting
thermal medium, various interactions take place and lead to 
dissipation: the jet loses some of its energy and sharpness.  
The latter phenomenon is referred to as jet broadening, or 
jet quenching. If its efficiency is measured experimentally as a function 
of the jet's energy (this can be done particularly well if the total
jet momentum is balanced against that of a hard photon, which does not
lose energy to the medium~\cite{exp}), then we may learn something about 
the properties of the medium itself. The current understanding is 
that in order to explain the jet quenching observed empirically 
in heavy ion collision experiments, interactions
have to be much stronger than suggested by leading-order 
perturbation theory 
(for reviews see, e.g., refs.~\cite{old}--\cite{new}).

On an intuitive level, a highly energetic jet can be thought 
of as a light-cone Wilson line, and the fact that we are probing its
fate in the transverse direction leads us to correlate the Wilson
line with a slightly 
displaced Hermitean conjugate. Adding lines at both ends  
leads to a light-cone Wilson loop. 
Arguments have been given to make the correspondence precise 
(see, e.g., refs.~\cite{phen1},\cite{def1}--\cite{def4}), 
however it appears difficult to state the 
form of the error that is made in this approximation.
In the following we take the light-cone Wilson loop as 
a starting point, without dwelling any further on its
relation to physically measurable quantities.  

In a statistical environment (with a temperature $T$, assumed to be above
a few hundred MeV), thermal noise leads to decoherence. As a result
the light-cone Wilson loop, to be denoted by $W$, ``decays''
at large Minkowskian times $t \gg \hbar/T$.\footnote{%
 Since the concept of a classical limit appears frequently, 
 it is useful to show $\hbar$ explicitly, thereby keeping the units of time 
 and energy separate. In contrast we set the speed of light equal to unity
 as usual.  
 } 
Schematically, assuming an appropriate time ordering, we may expect that 
\be
 \Bigl\langle W(t,r_\perp) \Bigr\rangle^{ }_T
  \; 
  \stackrel{ t \gg {\hbar} / {T} }{\sim}
  \; 
 Z(r_\perp)\, e^{-i V(r_\perp) t}
  \;\;
  \sim
  \;\;
 Z(r_\perp)\, e^{- i \rmi{Re} V (r_\perp) t} \, e^{-|\rmi{Im} V(r_\perp)| t} 
 \;, \la{sketch1}
\ee
where $r_\perp \equiv | \vec{r}_\perp |$ is the length of a 2-dimensional
transverse vector;
$\re V(r_\perp)$ is a real phase; 
and $\langle...\rangle^{ }_T$ refers to a thermal expectation value.  
If the coefficient of the exponential decay is represented 
in Fourier space, 
\be
 |\im V(r_\perp)| \; = \; \int_{\vec{k}_\perp}
 (1 - e^{i \vec{k}_\perp \cdot \vec{r}_\perp})
 \, C(k_\perp)
 \;, \la{sketch2} 
\ee
then $C(k_\perp)$ is often referred to as 
the ``transverse collision kernel''
(\cite{sch} and references therein). 
Considering for concreteness 
a Wilson loop in the fundamental representation, 
the leading-order expression for $C(k_\perp)$
at small transverse momenta reads~(\cite{agz}, \eq(44))
\be
 C(k_\perp)= g^2 T \CF 
 \biggl( \frac{1}{k_\perp^2} - \frac{1}{k_\perp^2 + \mE^2}
 \biggr) + \rmO\Bigl( \frac{g^4 T^2}{k_\perp^3} \Bigr)
 \;, \la{Ck}
\ee
where $g^2 \equiv 4 \pi \alpha_s / \hbar$ is the strong gauge coupling;
$\CF \equiv (\Nc^2 - 1)/(2\Nc)$; 
and 
\be
 \mE^2 \equiv \Bigl(\fr{\Nc}3 + \fr{\Nf}6 \Bigr) \frac{ g^2 T^2 }{\hbar} 
\ee
is the Debye mass parameter
(which has units of inverse distance squared). 
The question we are interested in is how
large the corrections to \eq\nr{Ck} can be, particularly within
the infrared domain $k_\perp \ll \pi T/\hbar$. 

Previous work already exists on infrared 
corrections to \eq\nr{Ck}. In particular, 
the corrections of $\rmO(g^4 T^2)$ 
were computed for $k_\perp\sim \mE$ in ref.~\cite{sch}, and non-perturbative
effects of $\rmO(g^6 T^3)$ for $k_\perp \sim g^2 T / \pi$ 
were addressed in ref.~\cite{nonpert}. In ref.~\cite{sch} it was noted
that for $k_\perp\sim \mE$ the perturbative series might be slowly 
convergent, and therefore in need of an all-orders resummation.
Conceptually, the aim of the current study is to implement such  
a resummation through numerical simulations of a low-energy description. 

More precisely, our goal is to address \eq\nr{sketch1}
within the framework of {\em classical lattice gauge theory} (CLGT). 
It should be immediately acknowledged that 
although CLGT does represent\footnote{%
 Originally CLGT simulations were employed for addressing 
 the rate of non-perturbative anomalous chirality violation 
 originating from the scale $k_\perp \sim g^2 T / \pi$, see e.g.\ 
 refs.~\cite{oldclas}--\cite{ew} and references therein. They have 
 also been used for studying the dynamics of thermal phase transitions, 
 see e.g.\ ref.~\cite{hr}, as well as many non-equilibrium problems
 in cosmology and heavy ion collision experiments.  
 } 
the physics of the system at scales $k_\perp \sim g^2 T / \pi$,
it actually is {\em not} quantitatively
accurate at the scales $k_\perp \sim \mE$ that are of most interest
here. The reason is that 
it is highly sensitive to lattice artifacts in this 
momentum range~\cite{bms,pa}. 
Nevertheless, 
it still contains the correct physics on the {\em qualitative} level; 
indeed CLGT simulations have been useful for gaining insight on various 
phenomena at the Debye scale (see e.g.\ refs.~\cite{imV,mink}), thereby
serving as a stepping stone towards full four-dimensional simulations
of the same problems 
(see e.g.\ refs.~\cite{ar}--\cite{mumbai}).
The great strength of CLGT is that it operates directly 
in Minkowskian space-time, thereby circumventing all issues
related to analytic continuation. The purpose of the present 
study is to explore what CLGT can teach us about the light-cone 
Wilson loop in the domain indicated in \eq\nr{sketch1}.\footnote{%
 Previously CLGT simulations have been used as an ingredient 
 in a phenomenological study of jet quenching of hard particles~\cite{bs}, 
 but the light-cone Wilson loop was not measured.
 }

The plan of this paper is the following. After outlining the general 
framework (\se\ref{se:frame}), we present some analytic
expectations in \se\ref{se:analytic}, setting the stage for a comparison
with numerical data. The numerical results are presented 
in \se\ref{se:numerics}, and we conclude in \se\ref{se:concl}.

%
\section{General framework}
\la{se:frame}

With a view on obtaining a formulation which may eventually be amenable 
to full four-dimensional lattice Monte Carlo simulations, we start by 
defining a ``tilted'' Wilson loop in Euclidean space-time. The Wilson loop 
is parametrized by a transverse extent, $r_\perp$; by an imaginary-time
variable, $\tau \in (0,\beta)$, where $\beta \equiv \hbar/T$; 
and by a velocity, $v_\rmii{E}$. At the 
end of the computation both $\tau$ and $v_\rmii{E}$ will be subjected
to a Wick rotation, but for the moment they are treated as real variables.  
The Wilson loop is illustrated in fig.~\ref{fig:loop}. In the limit 
$v_\rmii{E}\to 0$, it goes over into the Wilson loop defined  
in the context of heavy quarkonium physics in ref.~\cite{static}.
(We note that it may ultimately be more useful to ``tilt'' the 
thermal ensemble rather than the Wilson loop, which in four dimensions
can be achieved through shifted boundary conditions~\cite{gm1}.) 

%
\begin{figure}[t]

\hspace*{3cm}
\begin{picture}(200,100)(0,0)
\SetWidth{1.0}
\LongArrow(70,40)(115,10)%
\LongArrow(70,40)(70,100)%
\DashLine(70,40)(170,55){4}%
\LongArrow(130,49)(170,55)%
\DashLine(150,70)(150,28.5){1}%
\DashLine(100,20)(150,28.5){1}%
\SetWidth{2.0}
\Line(100,20)(150,70)%
\Line(100,20)(70,40)%
\Line(70,40)(115,85)%
\Line(115,85)(150,70)%
\Text(60,100)[c]{$\tau$}%
\Text(125,5)[c]{$r_\perp$}%
\Text(180,57)[c]{$r_\parallel$}%
\Text(165,26)[c]{$v_\rmii{E}\tau$}
\end{picture}

\caption[a]{\small 
An illustration of a tilted Wilson loop in Euclidean space-time. 
The slope in the ``parallel'' direction ($r_\parallel$) is parametrized by
a Euclidean velocity $v_\rmii{E}$, so that the right edge lies
at $r_\parallel = v_\rmii{E} \tau$. The transverse extent ($r_\perp$)
can be interpreted as the length of a two-dimensional vector.} 
\la{fig:loop}
\end{figure}
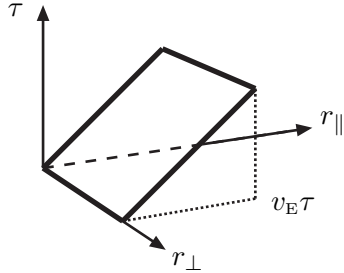
%

More concretely, starting with the continuum formulation and 
choosing sign conventions in which the covariant derivative
in the fundamental representation is
$D_\mu = \partial_\mu + i g_0 A_\mu$, a straight Wilson line reads 
\be
 W[X_2;X_1] = \mathcal{P}  
 \exp (-ig_0 \int_{X_1}^{X_2} \! {\rm d}X_\mu \, A_\mu)
 \;,
\ee
where $X\equiv (\tau,\vec{x})$. The foremost tilted line 
of \fig\ref{fig:loop} can be expressed as 
\ba
 W[(\tau,\vec{r}_\perp +  \vec{v}_\rmii{E} \tau);
    (0,\vec{r}_\perp)]
 & = &  
 \mathbbm{1} 
 - i g_0 
 \int_0^\tau \! {\rm d}\tau_1 \, (A_0 + \vec{v}_\rmii{E}\cdot\vec{A})
 (\tau_1,\vec{r}_\perp + \vec{v}_\rmii{E} \tau_1 )
 + \ldots \;,
\ea
where $\vec{v}_\rmii{E} \equiv v_\rmii{E}\, \vec{e}_\parallel$.
The expectation value of the Wilson loop is defined as 
\be
 C_\rmii{E}(\tau, v_\rmii{E}, r_\perp) \equiv 
 \frac{1}{\Nc}\tr
 \Bigl\langle
      W\bigl[(0,\vec{0}) ;\;
        (\tau,\vec{v}_\rmii{E} \tau) ;\;
        (\tau,\vec{r}_\perp +  \vec{v}_\rmii{E} \tau) ;\;
        (0,\vec{r}_\perp) ;\; 
        (0,\vec{0})\bigr]
 \Bigr\rangle^{ }_{T}
 \;, \la{CE}
\ee
where the thermal average 
$\langle ... \rangle^{ }_T$ 
implies periodic boundary conditions for bosonic and 
antiperiodic ones for fermionic fields over the Euclidean time direction. 
In the following, we have in mind evaluating the expectation
value within pure SU(3) gauge theory, even though this 
restriction can in principle be relaxed.  

Since the physical observable that we are interested in refers 
to Minkowskian time, an analytic continuation needs to be carried
out at the end of the computation. Technically, we do this by 
substituting $\tau \to i t$, which for 2-point functions yields
the time ordering corresponding to a Wightman correlator denoted
by $C^{ }_{>}$. (The Wilson loop can always be thought of as 
a 2-point function in time if the tilted lines are gauged to unity; 
general issues related to time ordering have been discussed 
in refs.~\cite{phen1},\cite{def1}--\cite{sch}.)  
However, since in the following we will 
simultaneously take the classical limit, 
time ordering actually plays no role. The classical limit
can be defined by writing 
\be
 \beta = \frac{\hbar}{T} \;, \quad g_0^2 = \ g^2 \hbar
 \;, \la{clas}
\ee 
and subsequently setting $\hbar \to 0$~\cite{db}. 
This limit is non-trivial and results in an interacting non-Abelian
gauge theory which captures the infrared features of 
the system's real-time thermal 
dynamics~\cite{oldclas,aaps}.

Apart from the continuum formulation, we also consider 
a lattice formulation of the theory in the following. Like in 
ref.~\cite{ks}, the theory
is discretized only in spatial directions, with a finite
lattice spacing $a$, whereas the time direction remains 
continuous.\footnote{%
 This formulation is invoked because of its close relation to CLGT;
 in contrast, the speculations to be made about full four-dimensional 
 lattice studies in \se\ref{se:concl} apply equally well to the 
 standard formulation with a symmetric discretization in all directions. 
 } 
Thereby the four-dimensional Euclidean action 
can formally be expressed as 
\ba
 S^{ }_\rmii{E} & \equiv & a^3 \sum_\vec{x} \int_0^{\beta} \! {\rm d}\tau \, 
 \Bigr\{ 
      \sum_{i=1}^{3} \tr [ E^2_i(X) ] 
   +  \frac{1}{a^4 g_0^2}\sum_{i, j = 1}^{3}
      \tr [\mathbbm{1} - P_{ij}(X)]
 \Bigr\}
 \;, \la{SE} 
\ea
where $g^{ }_0$ denotes the bare gauge coupling and 
$E_i$, $P_{ij}$ denote the electric field strength and the spatial 
plaquette, respectively: 
\ba
 E_i(X) & \equiv & 
 - \frac{i [ \partial_\tau U^{ }_i(X) ]U^\dagger_i(X) }{ag^{ }_0}
 + \frac{A_0(X) - U^{ }_i(X) A_0(X+a\vec{e}_{i}) U^\dagger_i(X)}{a}
 \;,  \\ 
 P_{ij}(X) & \equiv & U_i(X)\, U_j(X+a\vec{e}_{i})\, 
 U^\dagger_i(X+a\vec{e}_{j})\, U^\dagger_j(X)
 \;. 
\ea 
Here $U^{ }_i \in $~SU(3) are link matrices, and $A_0$ is a traceless
and Hermitean gauge field. The action is invariant 
under the gauge transformation
\ba
 U^{ }_i(X) & \to & G(X) U^{ }_i(X) G^{-1}(X + a \vec{e}_{i}) \;, \\ 
 A^{ }_0(X) & \to & G(X) A^{ }_0(X) G^{-1}(X) 
 + \frac{i}{g^{ }_0} [ \partial_\tau G(X) ] G^{-1}(X)
 \;,  
\ea
with $G \in$~SU(3). For perturbative computations we make use
of covariant gauges; in contrast, on the real-time simulation 
side it is convenient to make use of the corresponding Hamiltonian 
formulation with a vanishing Minkowskian $A_0$ and a corresponding
Gauss law constraint.

%
\section{Analytic expectations}
\la{se:analytic}

%
\subsection{HTL result in continuum}
\la{ss:continuum}

Our ultimate goal is 
to compute the analytic continuation of \eq\nr{CE} at large Minkowskian
times, $t \gsim \pi/(g^2T)$, and large transverse distances, 
$r_\perp \gsim 1/\mE$. We start, however, by inspecting
short distances, $r_\perp \lsim 1/\mE$.
This can be done with 
perturbation theory, provided that we recall that 
at high temperatures the 
loop expansion needs to be resummed to all orders 
in order to arrive at 
a consistent weak-coupling result. We are working at leading
non-trivial order in this regime, 
and then the effects of resummation are 
contained within Hard Thermal Loop (HTL)~\cite{htl1,htl2} propagators. 

Concretely, we carry out the computation by evaluating 
the graphs of \fig\ref{fig:graphs} with the Euclidean propagator
\be
 \Bigl\langle 
   A^a_\mu(X) \, A^b_\nu(Y) 
 \Bigr\rangle
 = 
 \delta^{ab}
 \Tint{K} e^{i K\cdot(X-Y)} 
 \biggl[ 
   \frac{\mathbbm{P}^\rmii{T}_{\mu\nu}(K)}{K^2 + \Pi^{ }_\rmii{T}}
  + \frac{\mathbbm{P}^\rmii{E}_{\mu\nu}(K)}{K^2 + \Pi^{ }_\rmii{E}} 
  + \frac{\xi K_\mu K_\nu}{K^4}
 \biggr]
 \;, \la{propA}
\ee
where $K \equiv (k_n,\vec{k})$ and 
$\xi$ is a gauge parameter. 
The projectors read
\be
 \mathbbm{P}^\rmii{T}_{\mu\nu}(K)  =  
 \delta^{ }_{\mu i} \delta^{ }_{\nu j}
 \biggl(
   \delta^{ }_{ij} - \frac{k_i k_j}{k^2} 
 \biggr)
 \;, \quad
 \mathbbm{P}^\rmii{E}_{\mu\nu}(K) = 
 \delta^{ }_{\mu\nu} - \frac{K_\mu K_\nu}{K^2} - 
 \mathbbm{P}^\rmii{T}_{\mu\nu}(K)
 \;.
\ee
The Euclidean propagators are expressed in a spectral representation, 
\be
 \frac{1}{K^2 + \Pi^{ }_{\rmii{T}(\rmii{E})} } = 
 \int_{-\infty}^{\infty} \! \frac{{\rm d}k_0}{\pi}
 \frac{\rho^{ }_{\rmii{T}(\rmii{E})}(\mathcal{K})}{k_0 - i k_n}
 \;, \la{spectral}
\ee
where $\mathcal{K} \equiv (k_0,\vec{k})$, and 
subsequently the Matsubara
sums are carried out. The explicit forms of the self-energies
can be found in the literature but are not needed here. 
In general the computation
parallels that in ref.~\cite{static}, except that it is in some sense
{\em simpler} (as long as we stay in continuum): 
indeed a non-zero $v_\rmii{E}$ ``regulates'' the contributions
of the Matsubara zero modes, so that they no longer need to be treated 
separately from the non-zero ones. 

%
\begin{figure}[t]
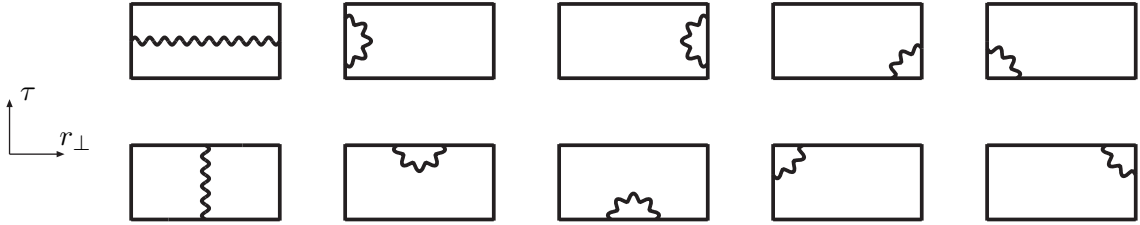


\begin{minipage}[c]{1.7cm}
$$
 \ScatLegend
$$
\end{minipage}
\begin{minipage}[c]{14cm}
\begin{eqnarray*}
&& 
 \hspace*{-2.6cm}
 \ScatF \qquad\quad 
 \ScatG \qquad\quad 
 \ScatH \qquad\quad 
 \ScatJ \qquad\quad 
 \ScatE \\[10mm] 
&& 
 \hspace*{-2.6cm}
 \ScatA \qquad\quad 
 \ScatB \qquad\quad 
 \ScatC \qquad\quad 
 \ScatD \qquad\quad 
 \ScatI
\end{eqnarray*}
\end{minipage}

\caption[a]{\small 
The graphs contributing to the tilted Wilson loop 
of \fig\ref{fig:loop} at $\rmO(g_0^2)$. Wiggly lines stand 
for HTL-resummed gluon propagators.} 
\la{fig:graphs}
\end{figure}
%

Some technical details of the computation are presented in appendix~A. 
Here we merely note that analytic continuation 
is carried out as $\tau\to i t$, 
$v_\rmii{E} \to -i v$, and the classical limit is taken as discussed around
\eq\nr{clas}.\footnote{%
 In practice the classical limit amounts to assuming that 
 $k_0 \ll \pi T / \hbar$; 
 therefore, at leading order it correctly represents the physics of 
 the large-time or low-energy limit of the exponential decay. 
 } 
The definition of a potential reads (cf.\ \eq\nr{sketch1})
\be
 i \partial_t C^{ }_\rmii{E}(it,-iv,r_\perp) \equiv 
 V(t,v,r_\perp) C^{ }_\rmii{E}(it,-iv,r_\perp)
 \;;
 \la{defV}
\ee
taking the limit $t\to \infty$ and setting $v\to 1$, we reproduce
the result of \eq\nr{Ck}:
\ba
 V^{(2)}_\rmi{cl}(\infty,1,r_\perp) & = &  
   -i g^2 T \CF 
  \int_{\vec{k}_\perp }
  \bigl(1 - \cos{\vec{k}_\perp \cdot \vec{r}_\perp} \bigr) 
  \biggl( \frac{1}{k_\perp^2} - \frac{1}{k_\perp^2 + \mE^2} \biggr)
 \la{Vexpl} \\ & = & 
  - i \, \frac{g^2 T \CF}{2\pi} 
  \biggl[ 
    \ln\Bigl( \frac{\mE r_\perp }{2} \Bigr) + \gammaE + K_0^{ }(\mE r_\perp)
  \biggr]
  \;. \la{Vexpl_r}
\ea
Here $K^{ }_0$ is a modified Bessel function.

Next-to-leading order (NLO) corrections to
the integrand of \eq\nr{Vexpl} have been 
determined in ref.~\cite{sch}. They are large and increase the 
magnitude of the imaginary part; their numerical 
contribution to \eq\nr{Vexpl_r} is shown 
in \fig\ref{fig:test} below.

%
\subsection{HTL result on a spatial lattice}
\la{ss:lattice}

For a practical measurement, the theory needs to be regularized; 
within CLGT, this means that we 
consider (a Minkowski-space classical limit of)
the theory defined by \eq\nr{SE}. Expressing everything
in lattice units and taking the limit of \eq\nr{clas}, 
the results depend on a single parameter, which we denote by 
\be
  \bG \equiv \frac{2\Nc}{g^2 T a}
  \;. \la{bG}
\ee
Initial configurations are generated with the weight 
$\exp(-\beta_\rmii{G} H_\rmi{cl})\prod_\vec{x}\delta(\mathcal{G}(\vec{x}))$, 
where
\be
 H_\rmi{cl} = \sum_\vec{x}
 \biggl\{
  \sum_{i=1}^{3} \tr [ \mathcal{E}^2_i(\vec{x}) ] 
  +
  \frac{1}{2\Nc}
  \sum_{i, j = 1}^{3}
      \tr [\mathbbm{1} - P_{ij}(\vec{x})] 
 \biggr\}
 \;; \la{Hcl} 
\ee
$\mathcal{G}(\vec{x})$ denotes the Gauss law constraint;
and $\mathcal{E}_i(\vec{x})$ are suitably normalized canonical momenta
conjugate to the link matrices $U_i(\vec{x})$. 
Subsequently the fields are evolved according to classical equations 
of motion (cf.\ \eqs\nr{dU}, \nr{dE}), 
and the observable is measured as illustrated in 
\fig\ref{fig:lattice}. 
(Further details on CLGT simulations 
can be found e.g.\ in refs.~\cite{aaps}--\cite{mota}, \cite{imV,mink}; 
the 
normalization of the electric field is strongly reference-dependent.)

%
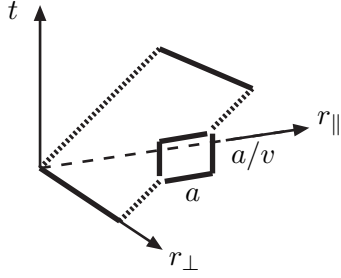
\begin{figure}[t]

\hspace*{3cm}
\begin{picture}(200,100)(0,0)
\SetWidth{1.0}
\LongArrow(70,40)(115,10)%
\LongArrow(70,40)(70,100)%
\DashLine(70,40)(170,55){4}%
\LongArrow(140,50.5)(170,55)%
\SetWidth{2.0}
\DashLine(100,20)(115,35){1}%
\DashLine(135,55)(150,70){1}%
\Line(135,38)(135,53)%
\Line(117,35)(135,38)%
\Line(115,37)(115,50)%
\Line(115,50)(133,53)%
\Line(100,20)(70,40)%
\DashLine(70,40)(115,85){1}%
\Line(115,85)(150,70)%
\Text(60,100)[c]{$t$}%
\Text(125,5)[c]{$r_\perp$}%
\Text(180,57)[c]{$r_\parallel$}%
\Text(128,30)[c]{$a$}%
\Text(151,45)[c]{$a/v$}%
\end{picture}

\caption[a]{\small 
An illustration of a tilted Wilson loop after the discretization
of the spatial directions with a lattice spacing $a$, and a Wick rotation
of both $\tau$ and $v_\rmii{E}$ to Minkowskian space-time. 
We represent the tilted Wilson lines by averaging the smallest
possible building blocks over the upper and lower paths. 
Measurements are taken at values $t = n a /v$, with 
$n \in \mathbbm{N}$. 
} 
\la{fig:lattice}
\end{figure}
%

Within CLGT, the Debye mass scale of the 
continuum formulation gets replaced with $\mE^2 \to g^2 T/a$, 
whereas the coupling constant scale remains put at $g^2 T$.
In lattice units, this implies that we want to determine the 
Wilson loop at separations $r_\perp /a \gsim \sqrt{\bG}$ and time
scales $t/a \gsim \bG$. The latter of these requirements poses
a significant challenge at large $\beta_\rmii{G}$, and introduces
a source of systematic errors with any limited resources. 
Approaching this regime from below, 
perturbation theory can again be used, but  
necessitates a HTL-type resummation, whose details were worked out
in refs.~\cite{bms,pa}. 

In practice, carrying out perturbative computations even to leading 
non-trivial order is cumbersome, due to the asymmetry in the
discretizations of the temporal and spatial directions. As an 
example, the expression obtained after carrying out the Wick 
contractions for the graphs in \fig\ref{fig:graphs}
is shown in appendix~B. As a main qualitative 
difference with respect to the continuum computation, we note
that the tilted Wilson lines in \fig\ref{fig:lattice} do {\em not} 
cancel against each other even at distance $r_\perp = 0$. Rather, 
we obtain an ``intercept'' which we denote by 
\be
 \mathcal{I}(v) \equiv 
 \frac{2 v a}{3}
 \int_{\vec{k}} \frac{\sin^2\Bigl( \frac{a \tilde k}{2v} \Bigr)}{\tilde k^2}
 \;, 
 \quad
 \tilde k \equiv \sqrt{\tilde k^2}
 \;, 
 \quad
 \tilde k^2 \equiv \sum_{i=1}^{3} \tilde{k}_i^2
 \;, \la{intercept}
\ee
where $\int_{\vec{k}}$ and 
$\tilde{k}_i$ are defined in \eq\nr{latt_defs}. 
Then we expect \eq\nr{Vexpl} to be replaced through 
\be
 V^{(2)}_\rmi{cl}(\infty,1,r_\perp) \simeq  
   -i g^2 T  \CF 
  \biggl\{ \mathcal{I}(1) +  
  \int_{\vec{k}_\perp} \bigl(1 - \cos{{k}_y {r}_\perp} \bigr) 
  \biggl( \frac{1}{\tilde{k}_y^2 + \tilde{k}_z^2}
   - \frac{1}{\tilde{k}_y^2 + \tilde{k}_z^2 + \mE^2} \biggr)
  \biggr\}
  \;, \la{Vexpl_lat}
\ee
where $\vec{k}_\perp \equiv (k_y,k_z)$, 
and the Debye mass parameter reads~\cite{bms,pa,math1,math2} 
\be
 \mE^2 = 2 g^2 T \Nc \frac{\Sigma}{4\pi a}
 \;, \quad
 \Sigma = \Gamma^2[\fr1{24}]\Gamma^2[\fr{11}{24}]
 \frac{\sqrt{3}-1}{ 48 \pi^2}
 \;. \la{mE_lat}
\ee
As discussed in appendix~B, 
the $r_\perp$-dependent part of \eq\nr{Vexpl_lat}
is an approximation, but is expected to be valid for
$r_\perp \gg a$. In any case \eq\nr{Vexpl_lat}
illustrates the general feature that, apart from the 
scale of the lattice spacing, the potential can have non-trivial structure
only at two distance scales, namely $1/g^2 T$ and $1/\mE$. 

%
\subsection{Beyond perturbation theory}
\la{ss:nonpert}

Let us extract lessons from above for what 
we may expect to see in the simulations: 
\bi
\item
At ``short'' distances, $r_\perp \ll 1/\mE$, $\im V_\rmi{cl}$
should start off with a non-zero intercept, given by \eq\nr{intercept} for
large $\bG$. 

\item
At ``intermediate'' distances, 
$r_\perp \sim 1/\mE$, the potential $\im V_\rmi{cl}$
shows a non-trivial structure which is relevant for jet quenching. 
This structure {\em cannot} be studied quantitatively with the 
approach of the present paper, given that within CLGT 
the Debye scale is completely determined by lattice artifacts, 
cf.\ \eq\nr{mE_lat}. On the qualitative level, however, we  
expect large corrections to the leading-order expression 
in \eq\nr{Vexpl_lat}~\cite{sch}.

\item
At ``long'' distances, 
$r_\perp \gg 1/\mE$, the phenomena related to the Debye 
scale are exponentially screened, and the physics is dominated by 
the colour-magnetic scale $g^2 T / \pi$.
More precisely, in continuum 
the imaginary part of the light-cone potential 
corresponds to the static 
potential of three-dimensional pure Yang-Mills theory~\cite{nonpert},
which for $r_\perp \gg \pi / (g^2T)$
evaluates to $|\im V_\rmi{cl}|  \simeq 0.553 (g^2 T)^2 r_\perp$ 
for $\Nc = 3$~\cite{mt}. When summed together with the NLO result
from ref.~\cite{sch}, which already includes a part of the linear 
term, the appropriate correction reads $\delta |\im V_\rmi{cl}|  \simeq 
[0.553 - {7}/({16\pi})] (g^2 T)^2 r_\perp$~\cite{nonpert}. 

\ei

\begin{figure}[t]

\centerline{%
\epsfysize=7.0cm\epsfbox{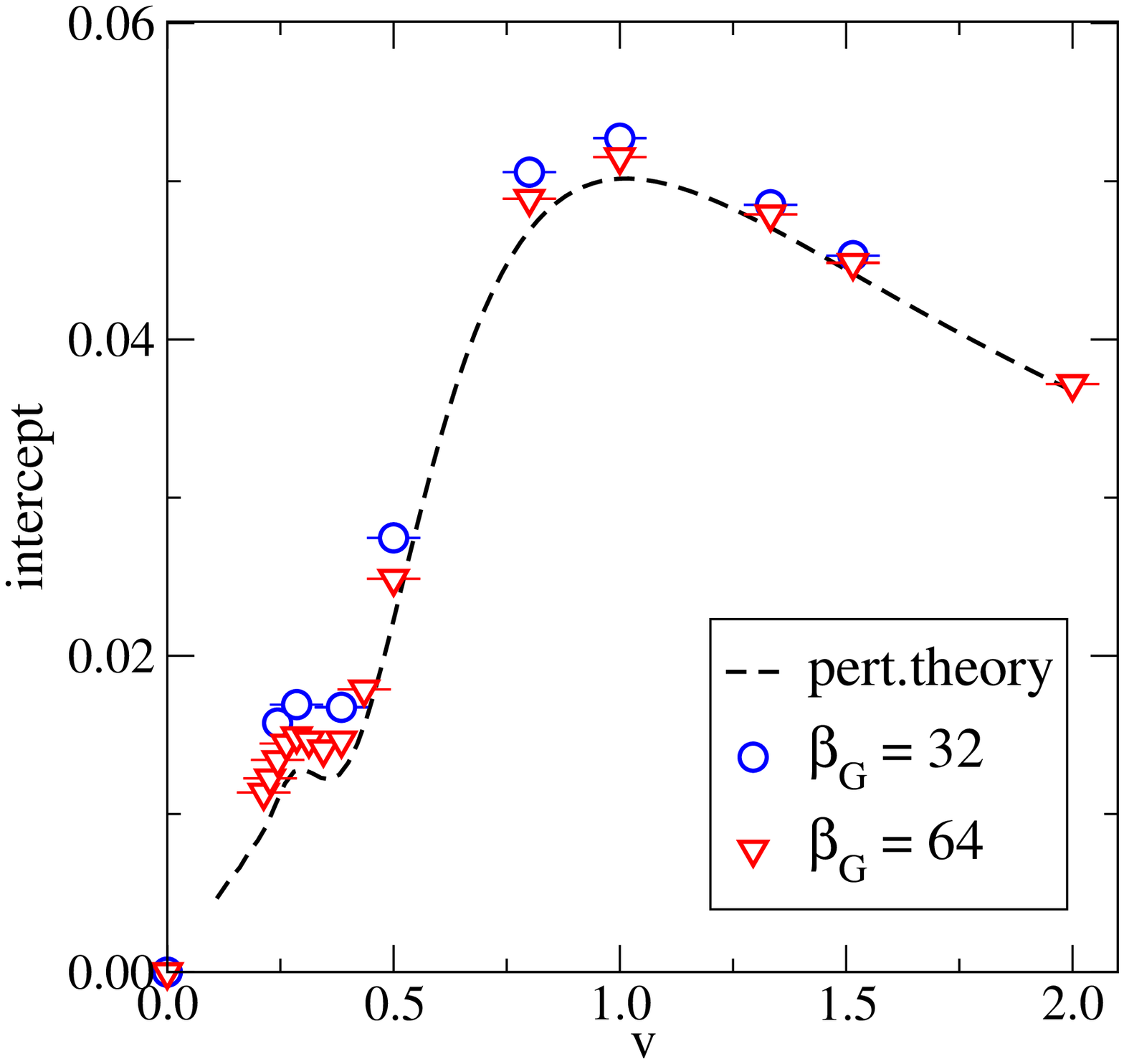}%
~~~\epsfysize=7.0cm\epsfbox{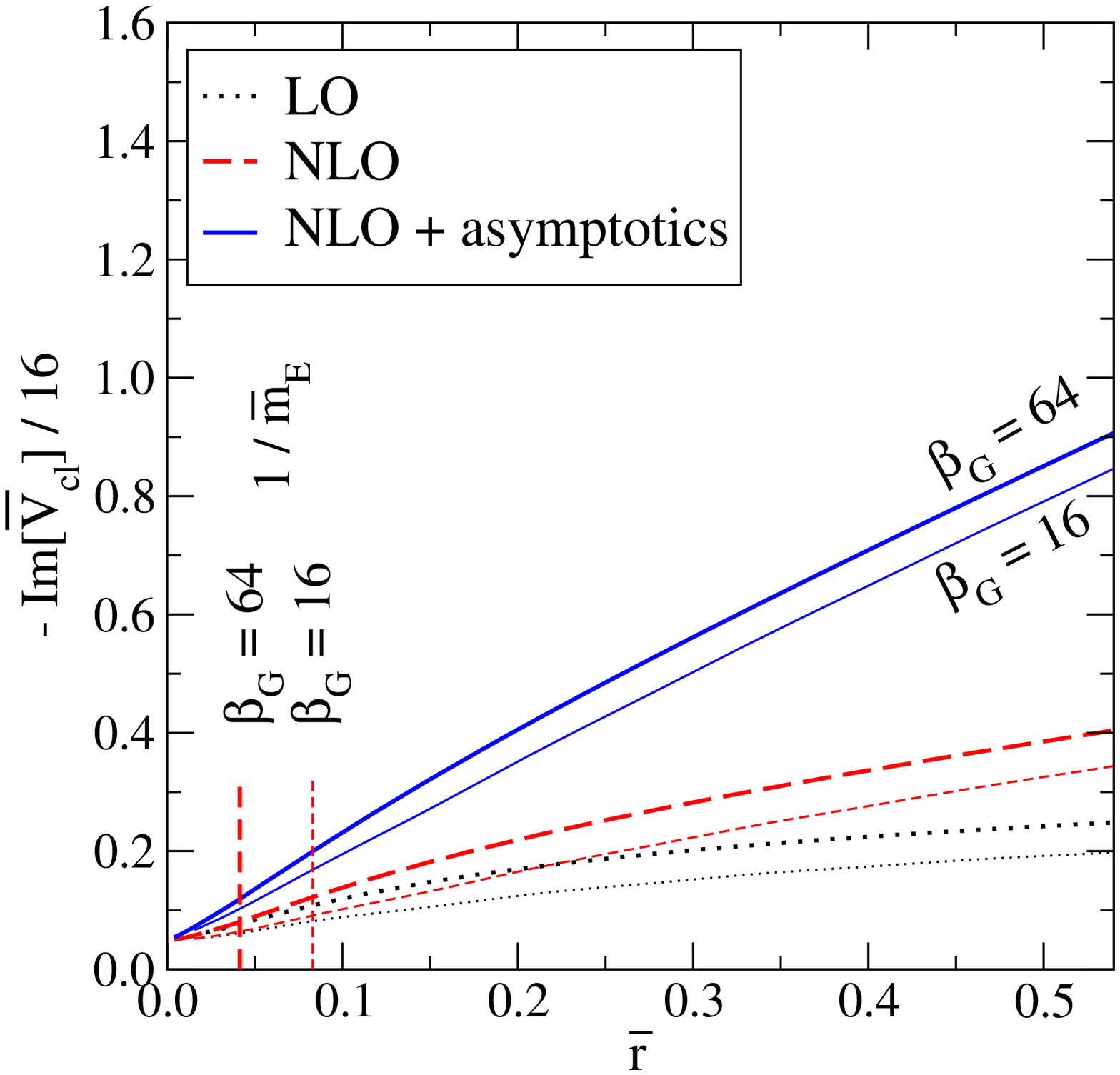}%
}

\caption[a]{\small Left: The intercept from \eq\nr{intercept}, in  
units of \eq\nr{icept_2}, compared with lattice data.
Right: The potential from \eq\nr{Vexpl_lat} (``LO''); 
with an integrand taken from ref.~\cite{sch} (``NLO''); and after adding 
the long-distance asymptotics~\cite{nonpert} (``NLO+asymptotics'').
Vertical lines indicate distances beyond which perturbation 
theory is unreliable. The asymptotic behaviour sets in 
at $\bar{r} \gsim 2 r_0 g^2 T /(2\Nc) \approx 0.73$.
Axis ranges have been chosen 
to agree with \fig\ref{MainResCLGT}(left) in which lattice data 
is shown. 
}

\la{fig:test}
\end{figure}

In order to observe the features mentioned 
in the data, it is helpful to change units. 
Suppose that we use $\bG$ from \eq\nr{bG} in order to convert 
lattice units to physical units. 
Then we can express distances and the potential as 
\ba
 \bar{r} & \equiv & \frac{r_\perp g^2 T}{2\Nc}
 \; =  \; \frac{r_\perp}{a\bG}
 \;, \la{r_units} \\
 \frac{\im \bar{V}_\rmi{cl}}{16} & \equiv &
 \fr1{16} \frac{2\Nc \im V_\rmi{cl}}{g^2 T}
 \; = \; \frac{\bG \im a V_\rmi{cl}}{16}
 \;, \la{pot_units}
\ea
where the factor 16 is a convention. 
In these units, the Debye scale corresponds to 
\be
 \frac{1}{\bar{m}_\rmii{E}} \; \equiv \; 
 \frac{g^2 T}{2 \Nc \mE} = \sqrt{\frac{\pi}{\beta_\rmii{G} \Sigma \Nc^2}}
 \;, 
\ee
the zero-distance intercept from \eq\nr{Vexpl_lat} amounts to 
\be
 \lim_{\bar{r} \; \ll \; 1/ \bar{m}_\rmii{E} } 
 \frac{ |\im \bar{V}_\rmi{cl}| }{16}  
 =  \frac{2 \Nc \CF \mathcal{I}(1) }{16}  = \frac{\mathcal{I}(1)}{2}
 \;, \la{icept_2}
\ee
whereas the long-distance asymptotics reads
\be
 \lim_{\bar{r} \; \gg \; 1 / \bar{m}_\rmii{E}} 
 \frac{ |\im \bar{V}_\rmi{cl}| }{16}  
 \approx \fr{1}{16} 0.553 (2\Nc)^2 \bar{r} 
 \approx 1.2 \bar{r}
 \;. 
\ee 
The various features together with the 
effect of NLO corrections~\cite{sch} are illustrated in \fig\ref{fig:test}.
The scale $1/\bar{m}_\rmii{E}$ defines the point beyond
which perturbation theory is no longer to be trusted, and the 
result may eventually 
(for $\bar{r} \gsim 1$) go over into the ``asymptotics'' curve reflecting
non-perturbative colour-magnetic dynamics. The non-perturbative
contribution of the scales $k_\perp \sim g^2 T/\pi$ to the so-called
jet quenching parameter, $\hat{q}$, is however determined 
by distances just above $r_\perp \sim 1/\mE$, 
rather than by the long-distance asymptotics~\cite{nonpert}.

%
\section{Numerical implementation}
\la{se:numerics}

\begin{figure}[t]

\centerline{%
\includegraphics[angle=-90,width=8.3cm]{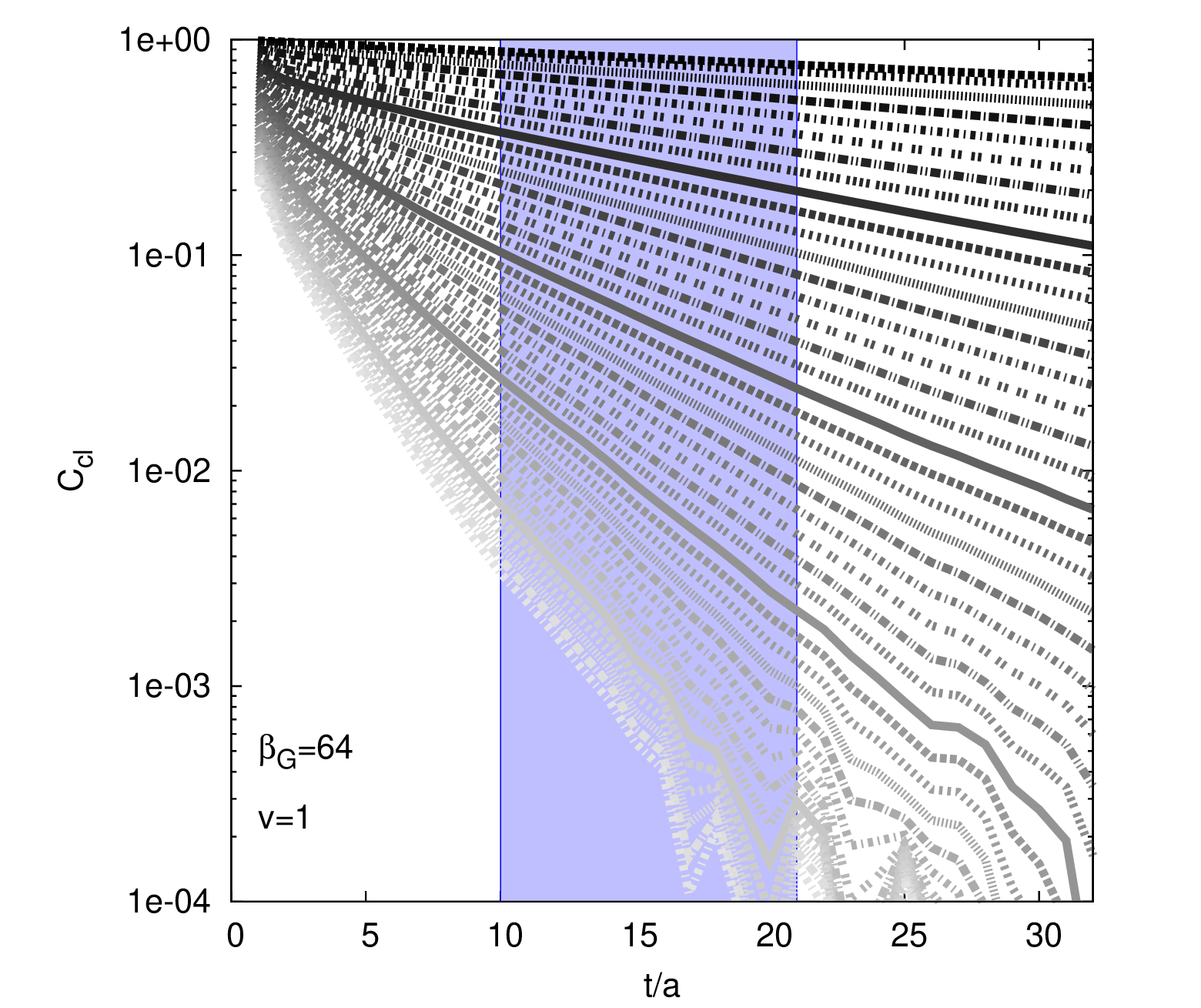}%
\includegraphics[angle=-90,width=7.3cm]{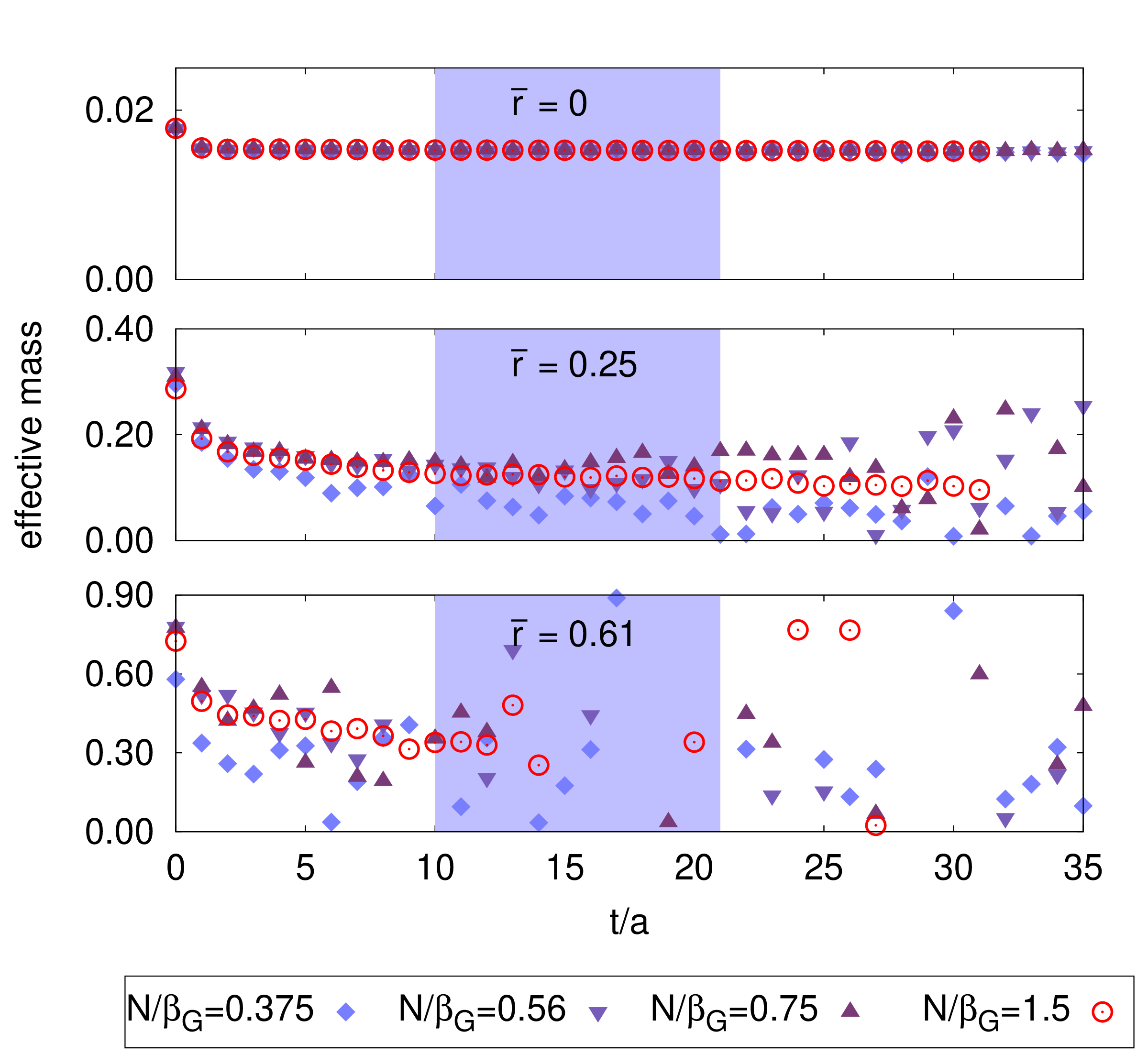}%
}

\caption[a]{\small
Left: Time dependence of the classical Wilson loop for $v=1$ 
at different distances~$\bar{r}$. The common fitting range for the 
determination of $\im V_\rmi{cl}$ is denoted by the shaded 
region. Right: Effective mass plots for 
three selected $\bar{r}$. The fitting range $t/a\in[10,20]$ 
is determined such that a satisfactory signal-to-noise ratio is 
obtained, however systematic errors could be substantial and the results
obtained should be thought of as upper bounds as usual. 
}

\la{WloopCLGT}
\end{figure}

\begin{figure}[t]

\centerline{%
\includegraphics[angle=-90,width=9.0cm]{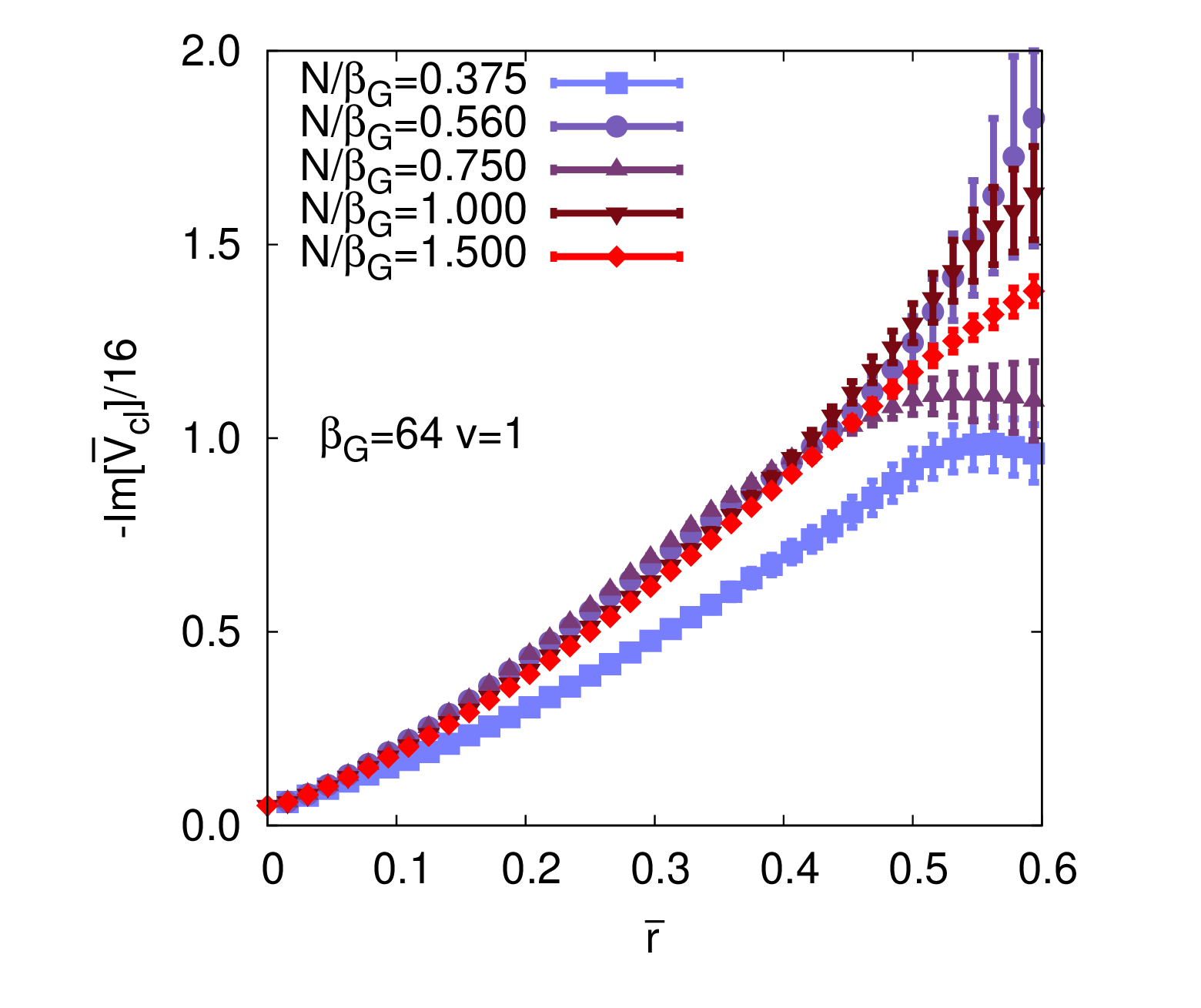}%
\hspace*{-1.5cm}%
\includegraphics[angle=-90,width=9.0cm]{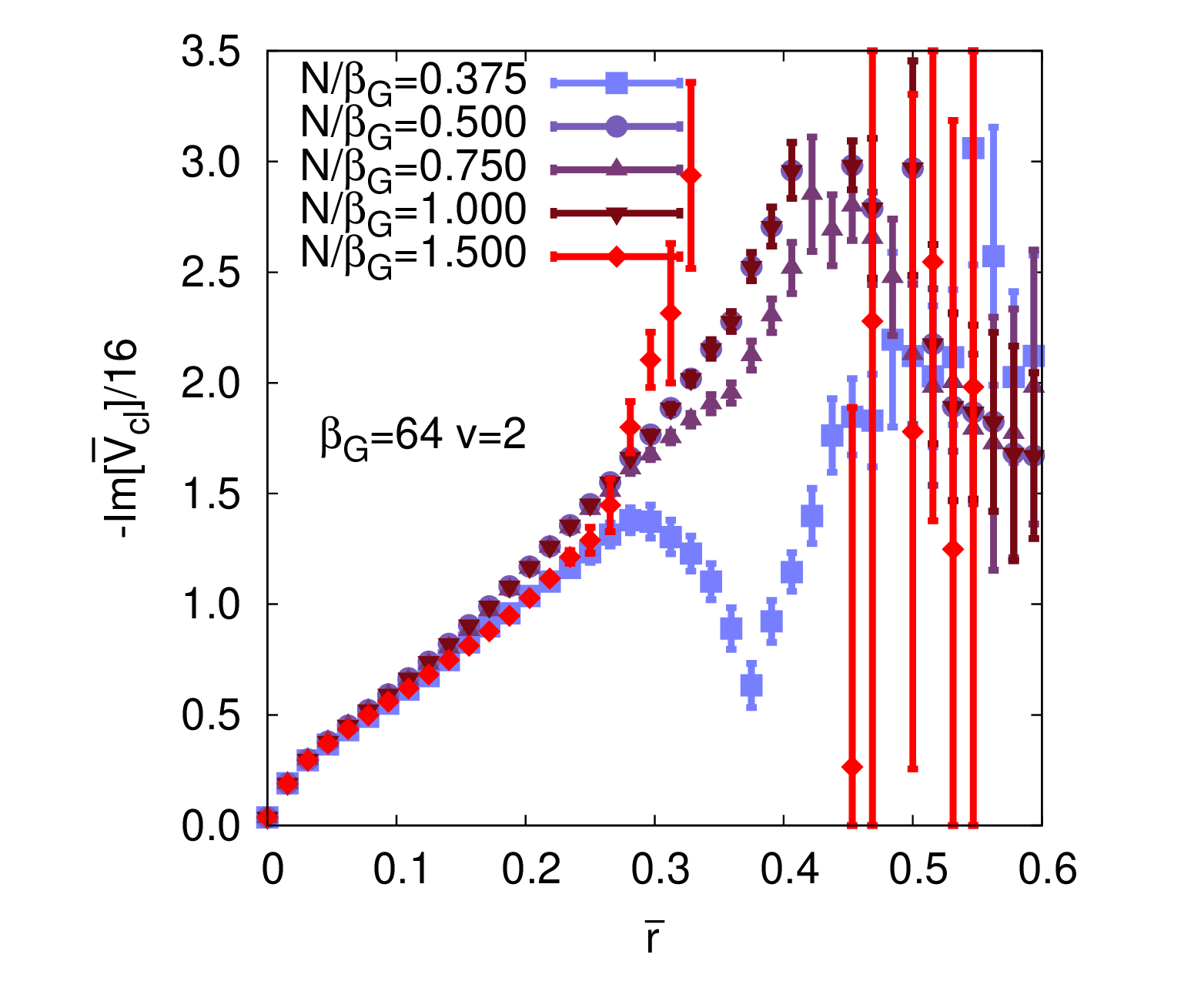}%
}

\caption[a]{\small
Volume dependence of the imaginary part of the potential 
as obtained from  lattice simulations with $\beta_\rmii{G}=64$, $v=1$ (left)
and $v=2$ (right), 
and a fitting procedure as described in \fig\ref{WloopCLGT}. 
}

\la{VolDepCLGT}
\end{figure}

\begin{figure}[t]

\centerline{%
\hspace*{0.5cm}%
\includegraphics[angle=-90,width=10.0cm]{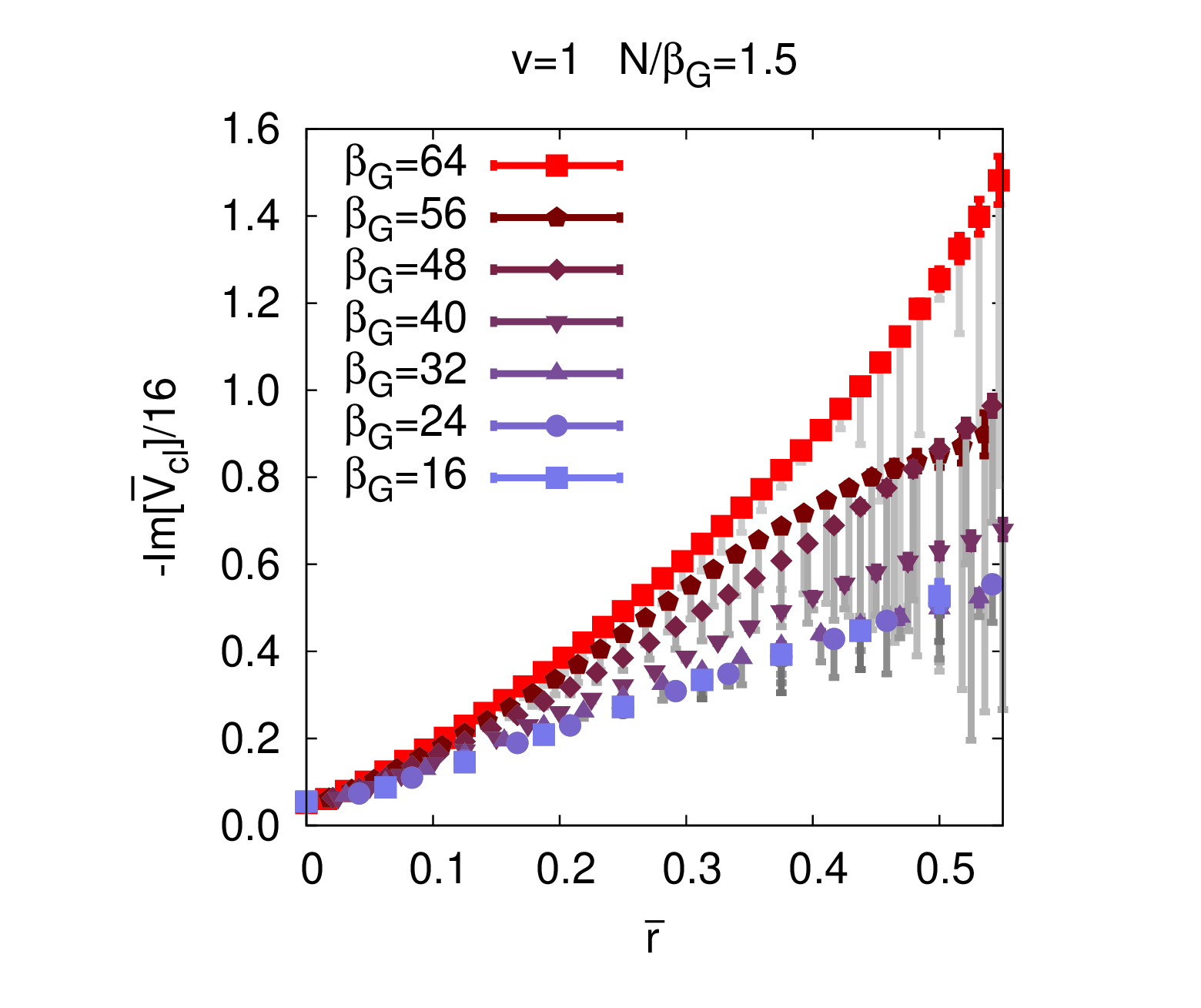}%
\hspace*{-2.0cm}%
\includegraphics[angle=-90,width=10.0cm]{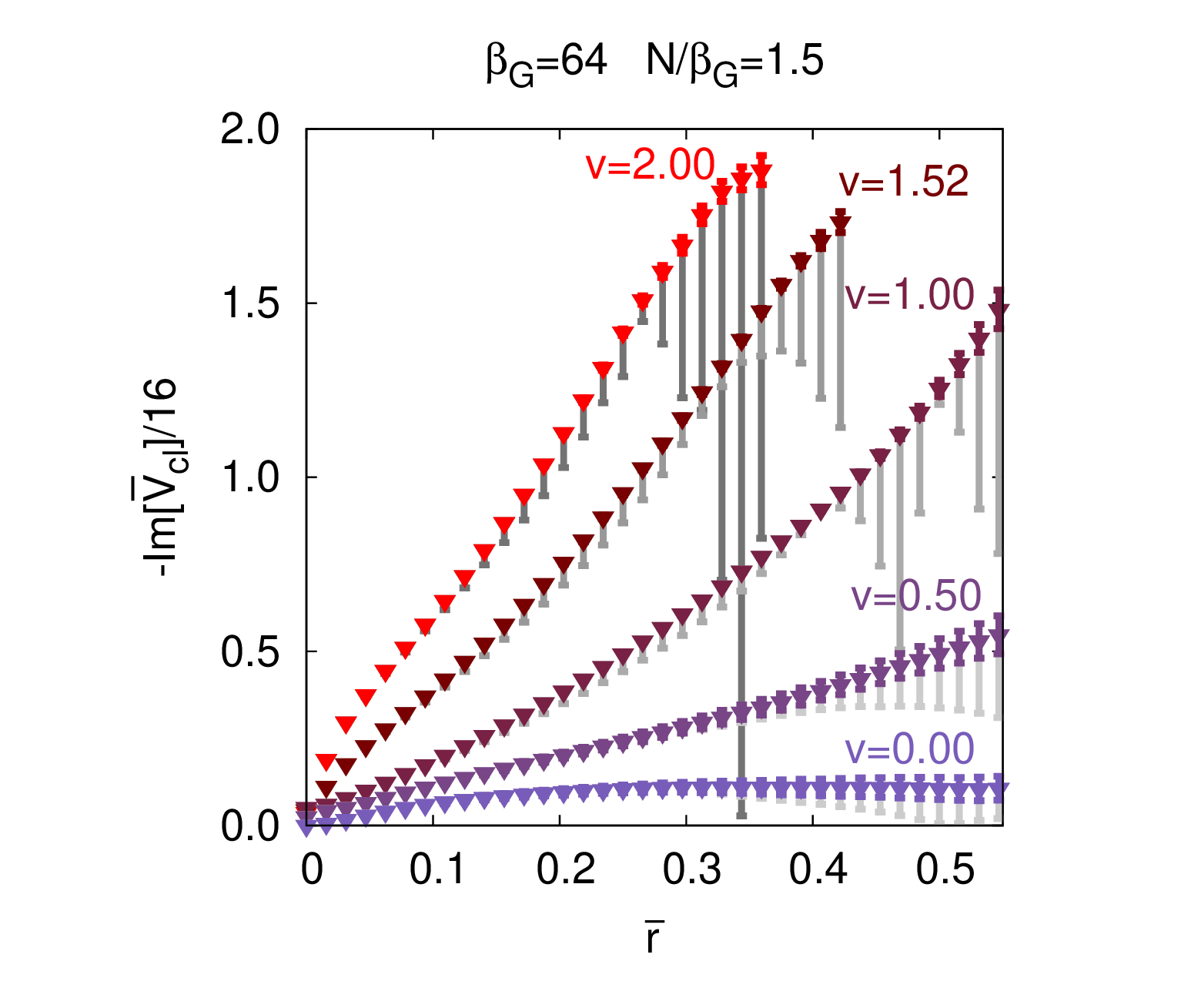}%
}

\caption[a]{\small
Left: 
The $\beta_\rmii{G}$ dependence of the imaginary part of the potential
for $v=1$ at a fixed physical volume $N / \beta_\rmii{G} =1.5$.
Right: Velocity dependence for
$\beta_\rmii{G} = 64$ and $N / \beta_\rmii{G} =1.5$. 
  Coloured bars denote statistical errors, whereas systematic errors, 
  estimated from pushing the fitting ranges to larger $t/a$, are
  given in gray (fitting at later times always decreases the result).
}

\la{MainResCLGT} \la{BetaDepCLGT}
\end{figure}

The simulations in the CLGT formalism are carried out along the lines of
ref.~\cite{imV}, in which the case $v=0$ was considered.  
With the choice of temporal gauge $U_0({\bf x},t)=1$ 
the classical equations of motion for the variables of \eq\nr{Hcl} read
\ba
 a\, \partial_t  U_i({\bf x},t) & = &  i \, (2 \Nc)^\frac{1}{2} 
 \mathcal{E}_i({\bf x},t) U_i({\bf x},t) 
 \;, \label{dU} \\
 a\, \partial_t \mathcal{E}^b_i({\bf x},t) & = &
 - \left( \frac{2}{\Nc} \right)^{\frac{1}{2}}
  {\rm Im} {\rm Tr} \Big[ T^b U_i({\bf x},t) \sum_{|j|\neq i} 
  S^\dagger_{ij}({\bf x},t) \Big] 
  \;, \label{dE}
\ea
where $S_{ij}$ denotes a staple. 
These differential equations are solved on a hypercubic
three-dimensional spatial lattice of size $N^3$
using the Euler forward finite-difference
scheme with temporal lattice spacing $a_t=\frac{a}{100}$. As initial
conditions, we deploy field configurations thermalized according to
refs.~\cite{aaps,ak,mr,imV} with the Hamiltonian of \eq\nr{Hcl} and the 
appropriate projection to the hypersurface respecting the Gauss law.

To obtain the potential of \eq\nr{defV}, 
we measure the discretized Wilson loop 
(defined like in \eq\nr{CE} but in Minkowski signature) 
in real time for several different transverse 
separations~$\bar{r}$. 
The tilting away from the temporal axis with velocity $v$ is
implemented as indicated in \fig\ref{fig:lattice}. 
Fitting the time evolution of
these purely real quantities (left panel of \fig\ref{WloopCLGT}) with an
exponential allows us to read off $\im V_\rmi{cl}$ from the exponent. 
For a rough estimate of $\lim_{t\to\infty}\im V_\rmi{cl}$, 
we identify a common fitting range for all values of $\bar{r}$, 
in which the asymptotic exponential falloff 
appears to have stabilized, while at the same time 
the statistical noise due to a finite number of measurements 
is still relatively small (right panel of \fig\ref{WloopCLGT}).
These requirements are hard to satisfy for large $\bar{r}$
and $\beta_\rmii{G}$ and, as can also be deduced from 
\fig\ref{WloopCLGT}(right) by bare eye, the procedure adopted
is likely to lead 
to an overestimate of  $\lim_{t\to\infty}\im V_\rmi{cl}$.

The effect of a finite volume on the determination 
of $\im V_\rmi{cl}$ is shown in
\fig\ref{VolDepCLGT} for $v=1$ and $v=2$. Higher velocities
lead to a faster exponential damping of the Wilson loop, hence the region for
an exponential fit shrinks and leads to a more noisy signal as shown
in the right panel. We find that to go to $\bar{r}\simeq 0.4$ 
a lattice extent of at least $N \gsim 1.5 \beta_\rmii{G}$ is necessary.

Once an adequate lattice extent and 
a usable fitting range  $t/a\in[10,20]$ have been established,
we proceed to measure the velocity dependence 
of $\im V_\rmi{cl}$ and
its intercept. For $\bar{r} \lsim 1/\bar{m}_\rmii{E}$
perturbation theory becomes more accurate 
at larger $\beta_\rmii{G}$, and indeed
the lattice results approach the perturbative ones for the intercept
at all velocities, cf.\  \fig\ref{fig:test}(left).

For $\bar{r} \gsim 1/\bar{m}_\rmii{E}$, in contrast, perturbation theory
need not be accurate. It is perhaps surprising then how well  
the ``NLO+asymptotics'' result works for moderate $\beta_\rmii{G}$, 
cf.\ \figs\ref{fig:test}(right) and \ref{MainResCLGT}(left), 
even though discrepancies remain at the 
smallest and largest $\beta_\rmii{G}$'s.
For the smallest $\beta_\rmii{G}$'s this may be due to the fact
that the Debye scale is larger [cf.\ \fig\ref{fig:test}(right)]
and therefore the asymptotics is approached at larger separations. 
For the largest $\beta_\rmii{G}$ we reiterate that 
it is difficult to reach the regime
$t/a \gsim \beta_\rmii{G}$ needed for extracting
the correct asymptotics (cf.\ \fig\ref{WloopCLGT}), so that 
the remaining discrepancy is probably due to systematic errors. 
Comparisons with perturbation theory need to be refined 
with other methods in the Euclidean domain~\cite{sch}, 
in which the Debye scale is free from lattice artifacts, so that 
the infinite volume and continuum limits can be systematically taken.  

%
\section{Conclusions and outlook}
\la{se:concl}

The purpose of this exploratory study has been to probe the contribution
that highly occupied 
classical gauge fields make to the thermal expectation
value of a light-cone Wilson loop. We have observed
that discrepancies to the leading-order
expression set in already at relatively 
short transverse distances, and lead to a larger magnitude 
of the imaginary part of the 
potential (stronger interactions) than predicted by
leading-order perturbation theory
(cf.\ \fig\ref{fig:test}(right) vs.\ \fig\ref{MainResCLGT}(left)). 
This is in qualitative agreement with the NLO computation 
of ref.~\cite{sch} and with the long-distance asymptotics 
as analyzed in ref.~\cite{nonpert}. 
Quantitative comparisons are hard because of 
discretization artifacts inherent to the CLGT framework. 

In addition, we have noted that crossing the light cone 
does not change the structure of the potential in any qualitative way 
(cf.\ \fig\ref{MainResCLGT}(right)). 
This poses well for the proposal 
of ref.~\cite{sch} according to which
the potential could be measured within a purely static dimensionally reduced
effective field theory~\cite{dr1,dr2}. 
Unlike classical lattice gauge theory, that 
framework is (super)renormalizable, so that divergences and discretization 
artifacts can be handled through local counterterms and 
analytic computations, and the genuine continuum physics 
of the momentum scale $k_\perp \sim \mE$ can be
disentangled. (It is useful to stress again 
that asymptotically large values of ${r}_\perp$ need {\em not}
be studied~\cite{nonpert}.) 
Thereby the existence of large infrared effects contributing to jet 
quenching can possibly be confirmed, perhaps leading to 
a QCD-based explanation for 
the experimentally observed efficient jet quenching
in current heavy ion collision experiments at the LHC.  

We would finally like to pose the question of whether the observable
of \eq\nr{CE} can also be addressed with direct four-dimensional
lattice simulations. 
One lesson from our study is that discretizing the tilted Wilson
lines (cf.\ \fig\ref{fig:lattice}) is inconvenient. 
It might rather be sensible to boost the ensemble by making
use of shifted boundary conditions~\cite{gm1}, and 
measure the Wilson loop always along the time-like lattice direction. 

Of course, measuring \eq\nr{CE} is not enough, but subsequently 
analytic continuations are needed for extracting the proper
real-time physics. In fact there are two 
separate analytic continuations here: $\tau \to i t$ as well as 
$v_\rmii{E} \to - i v$. The former is conventionally implemented 
by going through frequency space, i.e. estimating the spectral
function corresponding to the Euclidean correlator; from the spectral
function, any time ordering can be recovered. 
For $v_\rmii{E} = 0$, a determination of the spectral function {\em has}  
been attempted~\cite{ar,br,br2}, and even though systematic uncertainties
remain difficult to quantify, the challenge should not be much harder
in the presence of $v_\rmii{E} \neq 0$. Note that the quantity of interest here
corresponds to the {\em imaginary part} of the real-time potential, cf.\ 
\eqs\nr{sketch1}, \nr{sketch2}. 

As far as the analytic continuation of the velocity is concerned, 
one of the methods used in studies of QCD with 
a baryonic chemical potential might turn out to be helpful. 
For instance, one could first carry out simulations
with $v_\rmii{E}$; fit the results to a Taylor series; and subsequently
carry out an analytic continuation. Although in a mathematical sense
a singularity cannot be excluded as $v\to 1$, 
we have not observed any drastic changes 
in the {\em infrared dynamics} of the 
system in this limit
(the discretization-specific wobbles around $v\sim 0.3$
in \fig\ref{fig:test}(left) are 
not expected to be present if a boosted ensemble is simulated).
Therefore it is conceivable
that such a procedure could yield at least qualitative results against which
dimensionally reduced simulations, carried out on the space-like side 
of the light cone, can be compared.

%
\section*{Acknowledgements}

This work was partly supported by the Swiss National Science Foundation
(SNF) under grant 200021-140234 and by the European Commission under the FP7
programme HadronPhysics3. 

%
\appendix
\renewcommand{\thesection}{Appendix~\Alph{section}}
\renewcommand{\thesubsection}{\Alph{section}.\arabic{subsection}}
\renewcommand{\theequation}{\Alph{section}.\arabic{equation}}

%
\section{Leading-order perturbative computation in continuum}

We compute the graphs in  \fig\ref{fig:graphs} with the propagator
of \eq\nr{propA}, first in Euclidean space-time.  
Carrying out Wick contractions, it 
can be checked that any gauge parameter
dependence cancels. Inserting \eq\nr{spectral}
for $1/(K^2 + \Pi^{ }_\rmii{T(E)})$, the remaining expression reads
\ba
  C_\rmii{E}^{(0)}(\tau,v_\rmii{E},r_\perp) & = & 1
  \;, \\[2mm] 
  C_\rmii{E}^{(2)}(\tau,v_\rmii{E},r_\perp) & = & 
  \frac{ g_0^2 \CF}{\beta} 
  \int_\vec{k} \bigl(\cos{\vec{k}\cdot \vec{r}_\perp} -1 \bigr) 
  \int_{-\infty}^{\infty} \! \frac{{\rm d} \ko}{\pi}
  \sum_{k_n}
  \frac{2 - e^{i(k_n + \vec{k}\cdot \vec{v}_\rmii{E})\tau} 
  - e^{-i(k_n + \vec{k}\cdot \vec{v}_\rmii{E})\tau} }{ \ko - i k_n} 
  \nn & & 
  \times \biggl\{ 
    \rho_\rmii{E}(\ko,\vec{k})
    \biggl[ 
       \frac{1}{(k_n  + \vec{k}\cdot \vec{v}_\rmii{E})^2}
       \biggl( 1 + \frac{k_n^2}{k^2} \biggr)
    \biggr]
  \nn & & \;\; + \, 
  \rho_\rmii{T}(\ko,\vec{k})
    \biggl[ \frac{r_\perp^2}{(\vec{k}\cdot\vec{r}_\perp)^2} + 
       \frac{1}{(k_n  + \vec{k}\cdot \vec{v}_\rmii{E})^2}
       \biggl( v_\rmii{E}^2 - \frac{k_n^2}{k^2} \biggr)
    \biggr]
  \biggr\}
  \;. \la{CEtau}
\ea
Here $k_n \equiv 2\pi n/\beta$, with $n\in \mathbbm{Z}$, 
are the Matsubara frequencies. 
The apparent poles of \eq\nr{CEtau} 
at $k_n + \vec{k}\cdot\vec{v}_\rmii{E} = 0$
are regulated by the zeros of the numerator. 

The Matsubara sums can be carried out by partial fractioning 
the dependence on $k_n$, and then making use of 
\be
 \frac{1}{\beta} \sum_{k_n} \frac{ e^{i k_n \tau} }{\ko - i k_n}
 = \nB{}(\ko) e^{\tau \ko}
 \;, \quad
 \frac{1}{\beta} \sum_{k_n} \frac{ e^{- i k_n \tau} }{\ko - i k_n}
 = \nB{}(\ko) e^{(\beta - \tau) \ko} 
 \;, \quad
 0 < \tau < \beta
 \;, 
\ee
where 
$
 \nB{}(\ko) \equiv 1 / (e ^{\beta \ko} - 1)
$.
In order to simplify the expressions we also take the classical 
limit right away; recalling \eq\nr{clas}
and setting $\hbar\to 0$, the results then become
\ba
  C_\rmii{E,cl}^{(0)}(\tau,v_\rmii{E},r_\perp) & = & 1
  \;, \\[2mm] 
  C_\rmii{E,cl}^{(2)}(\tau,v_\rmii{E},r_\perp) & = & 
   g^2 T \CF 
  \int_\vec{k} \bigl(\cos{\vec{k}\cdot \vec{r}_\perp} -1 \bigr) 
  \int_{-\infty}^{\infty} \! \frac{{\rm d} \ko}{\pi}
  \frac{2 - e^{(\ko + i \vec{k}\cdot \vec{v}_\rmii{E})\tau} 
  - e^{-(\ko + i \vec{k}\cdot \vec{v}_\rmii{E})\tau} }{ \ko } 
  \nn & & 
  \times \biggl\{ 
    \rho_\rmii{E}(\ko,\vec{k})
    \biggl[ 
       - \frac{1}{(\ko  + i \vec{k}\cdot \vec{v}_\rmii{E})^2}
       \biggl( 1 - \frac{k_0^2}{k^2} \biggr)
    \biggr]
  \nn & & \;\; + \, 
  \rho_\rmii{T}(\ko,\vec{k})
    \biggl[ \frac{r_\perp^2}{(\vec{k}\cdot\vec{r}_\perp)^2} - 
       \frac{1}{(\ko  + i \vec{k}\cdot \vec{v}_\rmii{E})^2}
       \biggl( v_\rmii{E}^2 + \frac{k_0^2}{k^2} \biggr)
    \biggr]
  \biggr\}
  \;. \la{CEcl}
\ea
It can be observed that in the classical limit, the Matsubara sum
amounts effectively to replacing $k_n$ through $-i k_0$. 
(It would certainly be possible to keep 
$\nB{}(k_0)$ in an exact form, cf.\ ref.~\cite{nlo}
for $v^{ }_{ } = 0$, however only the Bose-enhanced
classical term $\nB{}(k_0) \approx T/(\hbar k_0)$ 
is expected to contribute to the large-$t$ limit
to be taken presently.)

Wick rotation is carried out through $\tau = i t$, $v_\rmii{E} = - i v$, 
and the potential is extracted from 
\be
 i \partial_t   C_\rmii{E,cl}^{(2)}(i t,-i v,r_\perp)
 \; \equiv \; 
 V^{(2)}_\rmii{cl}(t,v,r_\perp)
 \; C_\rmii{E,cl}^{(0)}(i t,-i v,r_\perp)
 \;. \la{eom}
\ee
We obtain
\ba
 V^{(2)}_\rmii{cl}(t,v,r_\perp) & = & 
   g^2 T \CF 
  \int_\vec{k} \bigl(\cos{\vec{k}\cdot \vec{r}_\perp} -1 \bigr) 
  \int_{-\infty}^{\infty} \! \frac{{\rm d} \ko}{\pi}
  \frac{e^{i (\ko + \vec{k}\cdot \vec{v}) t} 
  - e^{-i (\ko + \vec{k}\cdot \vec{v} ) t} }{ \ko } 
  \nn & & 
  \times \biggl\{ 
    \rho_\rmii{E}(\ko,\vec{k})
    \biggl[ 
       - \frac{1}{\ko  +  \vec{k}\cdot \vec{v} }
       \biggl( 1 - \frac{k_0^2}{k^2} \biggr)
    \biggr]
  \nn & & \;\; + \, 
  \rho_\rmii{T}(\ko,\vec{k})
    \biggl[ \frac{r_\perp^2 (\ko  +  \vec{k}\cdot \vec{v} )}
    {(\vec{k}\cdot\vec{r}_\perp)^2} + 
       \frac{1}{\ko  + \vec{k}\cdot \vec{v}}
       \biggl( v^2 - \frac{k_0^2}{k^2} \biggr)
    \biggr]
  \biggr\}
  \;. \la{Vcl}
\ea
Subsequently the large-time limit follows from 
\be
  \lim_{t\to \infty}
    \frac{e^{i (\ko + \vec{k}\cdot \vec{v}) t} 
  - e^{-i (\ko + \vec{k}\cdot \vec{v} ) t} }{ \ko + \vec{k}\cdot \vec{v} }
  = 2\pi i \, \delta(\ko + \vec{k}\cdot \vec{v})
  \;.  \la{limit}
\ee
Carrying out the integral over $\ko$ and setting also $v = 1$, 
so that $\vec{k}\cdot\vec{v} \to k_\parallel$, leads to
\ba
 V^{(2)}_\rmii{cl}(\infty,1,r_\perp) & = & 
   -i g^2 T \CF 
  \int_{\vec{k}_\perp} 
   \bigl(1 - \cos{\vec{k}_\perp \cdot \vec{r}_\perp} \bigr) 
  \nn & & \; \times \, 
  \int_{-\infty}^{\infty} \! \frac{{\rm d} k_\parallel}{\pi}
  \biggl\{ 
    \frac{ \rho_\rmii{T}(k_\parallel,\vec{k}) }{k_\parallel} 
   - \frac{ \rho_\rmii{E}(k_\parallel,\vec{k}) }{k_\parallel} 
  \biggr\} 
  \frac{k_\perp^2}{ k_\perp^2 + k_\parallel^2}
  \;. \la{Vasy}
\ea
Here we substituted $k_\parallel\to -k_\parallel$ for 
simplicity. This potential is purely imaginary and, according
to \eq\nr{eom}, corresponds to an exponential decay of 
the light-cone Wilson loop at large Minkowskian times, 
as anticipated by \eq\nr{sketch1}. 

The next step is to perform the integral over $k_\parallel$.
This is possible by re-expressing the spectral function as a 
discontinuity of the retarded correlator across the real axis, 
\be
 \rho(\ko,\vec{k}) = \frac{G_\rmii{R}(\ko+i0^+,\vec{k}) - 
 G_\rmii{R}(\ko-i0^+,\vec{k}) }{2i}
 \;, 
\ee
and by then carrying out the contour integral. In the literature
the procedure is known as a light-cone sum rule~\cite{agz}
(see also appendix~A of ref.~\cite{sch}), and yields 
\be
 \int_{-\infty}^{\infty}
 \! \frac{{\rm d}k_\parallel}{\pi}
 \frac{\rho(k_\parallel,\vec{k})}{k_\parallel} 
 \frac{k_\perp^2}{k_\perp^2 + k_\parallel^2}
 = G_\rmii{R}(0,\vec{k}_\perp)
 \;. \la{sumrule}
\ee 
The retarded 
propagator is, in turn, the analytic continuation of the Euclidean one. 

Recalling finally 
that the self-energy $\Pi^{ }_\rmii{T}$ of \eq\nr{propA} vanishes
at zero frequency, whereas $\Pi^{ }_\rmii{E}$ equals the Debye mass
parameter, $\mE^2$,  we recover \eq\nr{Vexpl}.

%
\section{Leading-order perturbative computation on a lattice}

If the computation of appendix~A is repeated in lattice
regularization, then the expressions
become a lot more complicated. For instance, 
employing the notation 
\be
 \tilde{k}_i \equiv \frac{2}{a} \sin \bigl( \frac{a k_i}{2} \bigr)
 \;, \quad 
 \undertilde{k_i} \equiv \cos \bigl( \frac{a k_i}{2} \bigr)
 \;, \quad 
 \int_\vec{k} \equiv \int_{-\pi/a}^{\pi/a} \!  
 \frac{{\rm d}^3\vec{k}}{(2\pi)^3}
 \;, \la{latt_defs}
\ee
and making use of Feynman rules derived from \eq\nr{SE}, 
the observable of \eq\nr{CEtau} can formally be expressed as 
($\vec{k} = (\vec{k}_\perp, k_\parallel$), 
$\vec{k}_\perp \equiv (k_y,k_z)$)
\ba
   C_\rmii{E}^{(2)}(\tau,v_\rmii{E},r_\perp) & = & 
  \frac{ g_0^2 \CF}{\beta} 
  \int_\vec{k} \bigl(\cos{{k}_y {r}_\perp}  -1 \bigr) 
  \int_{-\infty}^{\infty} \! \frac{{\rm d} \ko}{\pi}
  \sum_{k_n}
  \frac{2 - e^{i (k_n + {k}_\parallel {v}_\rmii{E} )\tau} 
  - e^{-i(k_n + {k}_\parallel {v}_\rmii{E})\tau} }{ \ko - i k_n} 
  \nn & & 
  \times \biggl\{ 
    \rho_\rmii{E}(\ko,\vec{k}) (\undertilde{k_\parallel})^2
    \Bigl(\widetilde{\scriptstyle \frac{k_n}{v_\rmiii{E}} } \Bigr)^2
    \biggl[ 
       \frac{1}{(\widetilde{\frac{k_n}{v_\rmiii{E}}  
             + k_\parallel })^2}
       \biggl( \frac{1}{k_n^2} + \frac{1}{\tilde{k}^2} \biggr)
    \biggr]
  \nn & & \;\; + \, 
  \rho_\rmii{T}(\ko,\vec{k})
    \biggl[ \frac{1}{\tilde{k}_y^2} + 
       \frac{1}{(\widetilde{\frac{k_n}{v_\rmiii{E}}  
             + k_\parallel })^2}
       \biggl(
         \Bigl(\undertilde{\scriptstyle \frac{k_n}{v_\rmiii{E}} } \Bigr)^2
         -
         \frac{1}{\tilde{k}^2} 
         \Bigl(\widetilde{\scriptstyle \frac{k_n}{v_\rmiii{E}} } \Bigr)^2
       \biggr)
    \biggr]
  \biggr\}
  \nn[3mm] & - & 
  \frac{ g_0^2 \CF}{\beta} \frac{\tau v_\rmii{E} a^3}{4} 
  \int_\vec{k}\int_{-\infty}^{\infty} \! \frac{{\rm d} \ko}{\pi}
  \sum_{k_n} \frac{1}{\ko - i k_n}
   \Bigl(\widetilde{\scriptstyle \frac{k_n}{v_\rmiii{E}} } \Bigr)^2
  \nn & & 
  \times \biggl\{ 
    \rho_\rmii{E}(\ko,\vec{k})
    (\widetilde{k_\parallel})^2
       \biggl( \frac{1}{k_n^2} + \frac{1}{\tilde{k}^2} \biggr)
  \nn & & \;\; + \, 
  \rho_\rmii{T}(\ko,\vec{k})
       \biggl(
         1
         -
         \frac{(\widetilde{k_\parallel})^2}{\tilde{k}^2} 
       \biggr)
  \biggr\}  
  \;. \la{CEtau_lat}
\ea
If we recall, however, that after the Matsubara sum and the classical
limit, $k_n$ gets essentially replaced by $-i k_0$, and that for non-zero
distances and large times the contribution emerges from 
$k_\perp \lsim \mE$ and $k_0, k_\parallel \lsim g^2 T/\pi$
(cf.\ \eq\nr{sumrule}), then the lattice four-momenta can to 
a good approximation be replaced by their continuum limits, 
\be
 \tilde{k}_\mu \to k_\mu
 \;,  \quad
 \undertilde{k_\mu}\to 1
 \;.
\ee 
Then the first structure of \eq\nr{CEtau_lat} goes over 
into \eq\nr{CEtau}. In contrast, the second structure, 
which is linear in $\tau$ and
independent of $r_\perp$, 
originates from self-energy corrections
of the tilted Wilson lines and 
is specific to lattice regularization. 
Since this short-distance contribution arises from ``hard'' scales, 
there is no need for resummation; we can replace the spectral
representations by free propagators, 
 \be
  \int_{-\infty}^{\infty} \! \frac{{\rm d}k_0}{\pi}
 \frac{\rho^{ }_{\rmii{T}(\rmii{E})}(\mathcal{K})}{k_0 - i k_n}
 \; \rightarrow \;
  \frac{1}{k_n^2 + \tilde{k}^2}
  \;.
 \ee
Subsequently the Matsubara sum, classical limit, and 
analytic continuation are taken as usual, which ultimately 
leads to the intercept of \eq\nr{intercept}.


\end{document}